\documentclass[conference]{IEEEtran}
\usepackage{blindtext, graphicx}

\usepackage{array}
\newcolumntype{M}[1]{>{\centering\arraybackslash}m{#1}}

\begin{document}

\title{Exploring the Use of RPAs as 5G Points of Presence}

\author{\IEEEauthorblockN{Javier Suarez\IEEEauthorrefmark{1}, Ivan Vidal\IEEEauthorrefmark{1}, Jaime Garcia-Reinoso\IEEEauthorrefmark{1}, Francisco Valera\IEEEauthorrefmark{1}, Arturo Azcorra\IEEEauthorrefmark{1}\IEEEauthorrefmark{2}}
\IEEEauthorblockA{\IEEEauthorrefmark{1}Universidad Carlos III de Madrid -  Avda. de la Universidad 30, 28911 - Leganes, Madrid (Spain)}

\IEEEauthorblockA{\IEEEauthorrefmark{2}IMDEA Networks - Avda. Mar Mediterraneo 22, 28918 - Leganes, Madrid (Spain)
}
}

\maketitle

\begin{abstract}
This paper presents an early exploration and preliminary results on the use of Remotely Piloted Aircrafts (RPA) as 5G points of presence. The use of RPAs in the 5G arena would enable a cost-effective deployment of functions over mobile nodes that could be integrated on demand into the programmable and unified 5G infrastructure, enhancing the capacity of the network to flexibly adapt to the particular service requirements in a geographical area. As a first step, we evaluate the feasibility and the cost, in terms of energy consumption, of using virtualisation techniques over resource-constrained aerial vehicle platforms, as a fundamental software technology in the evolution towards 5G. We complement this evaluation presenting a proof-of-concept that considers the use of these platforms to enable real-time 5G communications in emergency cases.

\end{abstract}




\section{Introduction}
The eagerness to satisfy the demands imposed by their users has boosted telecommunications stakeholders to make incremental advances in communication networks. However, just after the deployment of a new network generation, new services shortly arise demanding new features to such networks. Thus, it is extremely important to
define a new network architecture focused on flexibility, which is the aim of 5G networks. In order to achieve such flexibility, 5G networks will be driven by software, reducing 
the requirement of expensive dedicated hardware. Because of this softwarisation, new trending technologies and concepts like Software-Defined Networking (SDN), 
Network-Function Virtualisation (NFV), Mobile Edge Computing (MEC) and Fog Computing (FC) play a crucial role in 5G. This will evolve  current rigid networks towards programable environments. 

Although flexibility is a requirement in 5G networks, the resulting architecture should satisfy an ambitious set of key performance indicators (KPI) \cite{networld2020}. These KPIs go from the increment of the network capacity to connect more devices (10 to 100 times more than the current generation), the reduction of the end-to-end latency (less than 1ms) to the reduction of service deployment time. To tackle these, and other KPIs, it is necessary to design and develop new cutting-edge technologies like small and pico cells, on-the-edge caches, ubiquitous virtualisation platforms, etc. Furthermore, 5G networks must be sustainable too, reducing CapEx and OpEX, which is challenging taking into consideration the remaining KPIs. For example, to increase the network connectivity in the downtown of a given city, one infrastructure operator could deploy a dense network of small cells connected to macro cells. Although this infrastructure could accomplish the KPIs defined in \cite{networld2020} during the peak hours, its expensive resources will be underused during other periods like night or the weekend.

In this paper, we introduce a preliminary study that explores the viability of developing 5G nodes on Remotely Piloted Aircrafts (RPA), and in particular on the more resource-cosntrained Micro Aerial Vehicles (MAV).  With such type of devices, an infrastructure provider has the flexibility to move resources on-demand, from one area to those where is needed. A main characteristic of MAVs is their ability to move and position on specific locations, although there are other functionalities that can be provided by them: the possibility to transport different types of radio links (WiFi, LTE, etc.), providing caching near the end-user, capacity to deploy virtualised services as well as virtualised network functions on the edge, etc. These characteristics may help infrastructure providers to accomplish several KPIs in an cost-effective way. Nevertheless, MAVs are resource-constrained devices, where their reduced size, payload transport capacity and reduced battery life may impose important restrictions to the previous functionalities described here. As a first step in our work, in this paper we focus on how to deploy network and service virtualisation on resource-constrained MAV platforms. 

The rest of the paper is organised as follows. Section~\ref{sec:rpas} presents a related work about remotely piloted aircraft systems. Section~\ref{sec:use-cases} describes the benefits of using MAVs as 5G points of presence, providing a set of use cases to highlight those benefits. Section~\ref{sec:preliminary-results} shows some preliminary results that validate the viability of using virtual machines to deploy 5G functions on resource-constrained devices. Finally, Section~\ref{sec:conclusion} concludes the paper and presents our future work.

\section{Remotely Piloted Aircraft Systems}
\label{sec:rpas}
Nowadays Europe is facing major societal challenges that require the simultaneous usage of resources and knowledge related to various heterogeneous fields. Among these, a key challenge is to use new information and communications technologies to build secure societies. This challenge encompasses the research and innovation activities needed to enhance the capacity of the society to accommodate extreme situations caused by natural or man-made disasters, to fight against crime and terrorism or to improve surveillance and border security operations, among others. In this context, the development of Remotely Piloted Aircraft System (RPAS) technologies is acquiring a fundamental relevance. 

An RPAS typically includes a set of Remotely Piloted Aircraft (RPA) units subordinated to a ground control station, which coordinates the execution of the mission-oriented application that led to the RPAS deployment. To support the mission objetives, each RPA may transport different payloads, such as communication equipment, daylight/thermal video cameras or diverse sensor systems. Although RPAS were initially developed to support military operations, they are currently being considered as enablers of mission-oriented civilian applications. In this respect, the emergent micro aerial vehicle (MAV) platforms are obtaining an increasing interest from the research community and the industry. With reduced cost and power consumption, compared with larger RPAS, these small-sized drones open new and exciting possibilities to execute collaborative applications \cite{quaritsch2008}, such as cooperative search, the collaborative generation of images in emergency situation, setting up aerial sensor networks to aid in disaster management or even structure building. Probably the main reason preventing MAVs from being massively deployed, is the complex regulation that is being required in most countries so to be able to guarantee safe flight conditions on non-segregated (civilian) airspace. Anyway, a change in regulations should be expected for specific services like those described in this work. 

As related work for this paper, there are several examples in the literature that consider the use of RPAS as an underlying platform to provide networking, computing and storage resources. The support of communications through unmanned airborne vehicle relays has been long considered \cite{pinkney1996} \cite{burdakov2010}. In \cite{palat2005} and \cite{rubin2007} they elaborate on this idea, proposing the use of RPAS as communication relays for ad-hoc ground networks. Similar approaches use RPAS to aid reliability of data communications in wireless sensor networks \cite{pignaton2010}. In \cite{charlesworth2013}, they present a communication scheme where two RPAS execute an algorithm based on game theory to relocate and maximize their joint coverage to a community of ground mobiles. In \cite{vidal2014} the authors describe a 
communication system for RPAS, which uses the computation capabilities of the RPA to select the most appropriate data-link for the communications between the aircraft and the ground control station. The work in \cite{wang2007} addresses cooperative formation flying with obstacle avoidance, using a technique with computational and storage requirements such that it can be implemented in limited-capacity aircrafts. In \cite{jawhar2015}, the authors propose severals strategies to support data collection in wireless sensor networks, based on unmanned aerial vehicles. The work in \cite{peng2013} describes a specific strategy to address data collection in this scenario, using a set of cooperative RPA units.

\section{Using RPAS as 5G points of presence}
\label{sec:use-cases}
As indicated by the related work, RPA platforms may provide diverse networking, computing and storage resources, which may be used for heterogeneous purposes and applications. Motivated by this observation, we argue that RPAS may represent an adequate platform to execute 5G functions based on the key software technologies envisioned for this new network paradigm (SDN, NFV, MEC and FC). 

\begin{figure*}[!t]
\centering\includegraphics[width=0.7\textwidth]{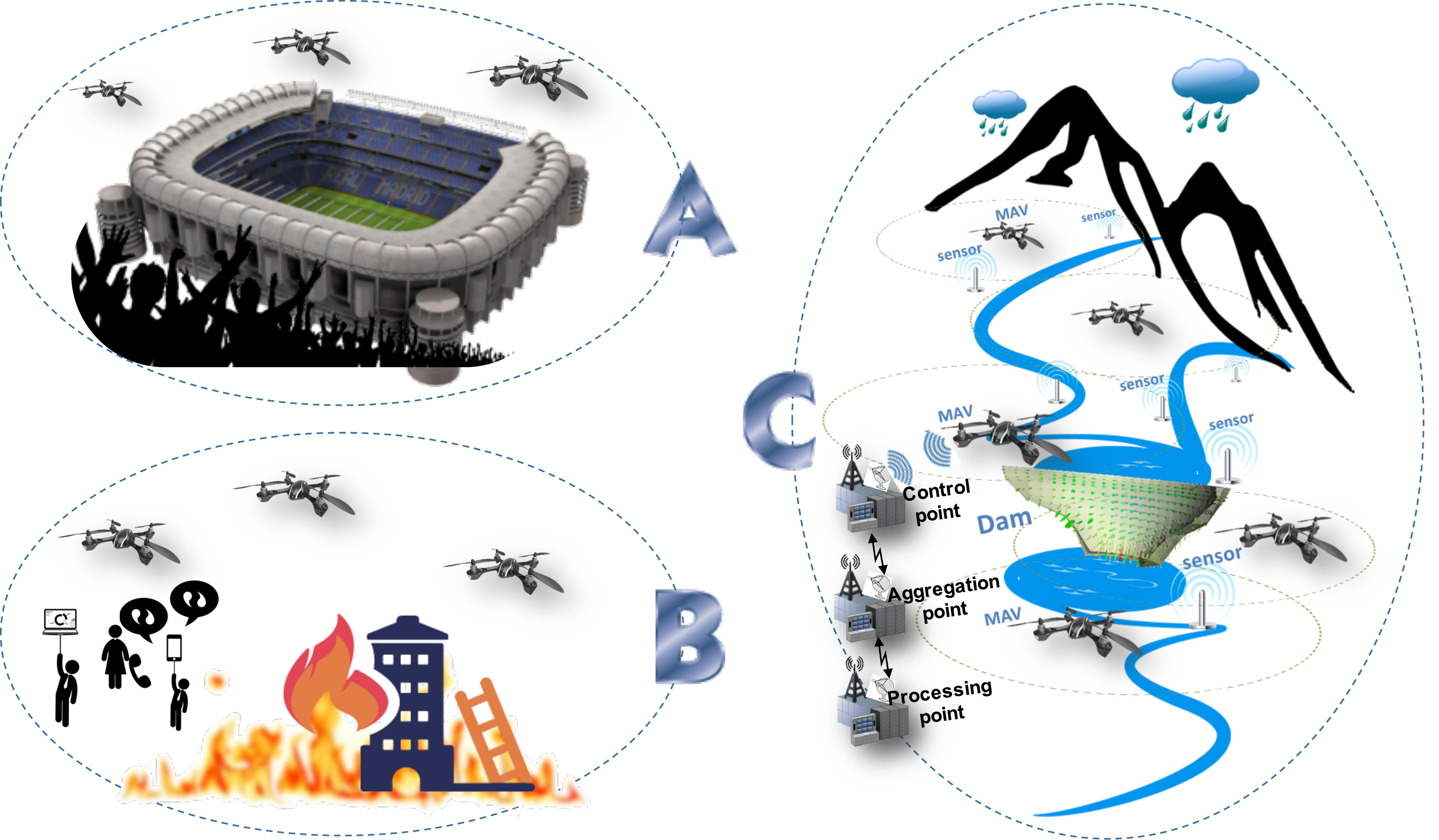}
\caption{Use cases}
\label{fig:use-cases}
\end{figure*}

Table \ref{table:KPIs} shows a subset of the Key Performance Indicators (KPIs) \cite{networld2020} defined for 5G, which are specially relevant in the scope of this paper. In the following, we showcase the advantages of using RPA units to deploy 5G functions with a set of illustrative use cases, making reference to the aforementioned KPIs where appropriate. These use cases are represented in Fig.~\ref{fig:use-cases}.
\begin{table}[ht]

\caption{5G Key Performance Indicators (KPIs)}
\begin{tabular}{|>{\centering\arraybackslash}m{3cm}|>{\arraybackslash}m{5cm}|}
\hline
\textbf{Throughput (KPI 1)} & 1000x more available aggregate throughput, 10x for individual users \\
\hline
\textbf{Latency (KPI 2)} & Service-level latency down to 1ms (when needed) \\
\hline
\textbf{Service creation time (KPI 3)} & From the application down to the network level, in the order of seconds or less \\
\hline
\textbf{Coverage (KPI 4)} & Seamless extension of 5G services anywhere anytime \\
\hline
\textbf{Total Costs of Ownership (KPI 5)} & Sustainable in terms of revenue generation and investments \\
\hline
\end{tabular}
\label{table:KPIs}
\end{table}

\subsection{User experience continuity in dense areas}
RPAs may provide a cost-effective solution to adapt the network infrastructure to increased service demands from a specific geographical area. As an example, consider the case of a sporting event or a concert taking place in a stadium. This creates the challenge to support high data rates and reduced latencies to ensure user experience continuity, as tens of thousands of user devices will demand network connectivity to establish voice/video conversations, send text messages, exchange images and video through social networks or even broadcast the event in high definition. In this situation, an infrastructure provider could deploy a number of MAVs over the stadium, supporting different network access technologies (e.g., WiFi, 3G/4G or even novel 5G radio technologies), to complement the network infrastructure that may be already available in the area of coverage. The trajectories and position of the MAVs would be governed from a ground control station, allowing the rapid establishment of the network infrastructure (KPI 3) before the event starts. MAVs would deploy 5G functions, providing a programmable infrastructure to offer networking, processing and storage resources over the deployment area, which will be available to operators/providers that provide network services to the users. This would allow operators to adapt their network infrastructure to throughput and latency demands (KPIs 1 and 2). This MAV infrastructure could be retired after the event, and be utilised to provide similar services in other dense areas as required. This allows amortising the investment costs and facilitate the sustainability of the infrastructure (KPI 5).

\subsection{Service provisioning in emergency situations}
The use of RPA platforms allow supporting the fast and on-demand deployment of aerial vehicles over delimited geographical areas. This may be of particular usefulness to provide network services in emergency situations, such as fire extinction in remote locations or to aid search and rescue operations in disaster areas (in both cases, communication services may be non-existent, insufficient or unavailable). In such cases, an infrastructure of MAVs could be rapidly deployed to integrate 5G points of presence within the disaster area (KPI 4). This would allow conforming an appropriate network/service infrastructure in a reduced timeframe (KPI 3). This infrastructure would enable real-time voice/video communications among the different units of the emergency response team, providing not only the network resources needed to deliver the media with the required latency constraints (KPI 2), but also the network elements that are needed to support the execution of signalling procedures between user terminals (e.g., SIP call session control functions). Unmanned ground vehicles and separate MAVs could also behave as users of the infrastructure, utilising it to deliver real-time video and sensed data that might be relevant to the emergency situation. Larger RPA units could be utilised to support long-distance data communications towards control centres responsible of coordinating the emergence operations. 

\subsection{Aiding IoT coverage to wide areas}
The Internet of Things (IoT) is expected to be a widely adopted technology in the near future. 5G must embrace this new network paradigm by providing network coverage anywhere, anytime and to anything. In this respect, RPAs may acquire a special relevance, as a cost-effective platform to enable non real-time communications with IoT devices. As a specific example, consider an IoT deployment where multiple sensors are installed in a geographical area to collect relevant hydrological data, such as information on the volume of flow circulating through rivers and tributaries, the level of water in dams, or meteorological information collected in riverbeds. In this scenario, a swarm of MAVs could provide the networking and storage resources necessary to collect and deliver the hydrological information to a set of control points, which would transmit these data towards a processing center through aggregation points (this example corresponds to a realistic communication topology used in the Spanish basins\footnote{SAIH: Automatic Hydrologic Data Collection System. http://www.magrama.gob.es/en/agua/temas/evaluacion-de-los-recursos-hidricos/SAIH/default.aspx}). This way, the use of RPAs allows the collection of hydrological information from remote locations (KPI 4), avoiding the costs associated with the deployment of additional control points and with satellite communication (as this may be the only communication technology available in remote geographical areas typically considered in this type of scenarios). As another illustrative example, RPA platforms could also be used to complement the existing network infrastructure in urban environments to support data communications in IoT deployments (e.g., to deliver data periodically generated by environmental sensors).

\section{Preliminary results}
\label{sec:preliminary-results}
As a first step towards exploring the use of MAVs as 5G point of presence, in this section we present some preliminary results related with deploying virtualisation environments (as a key technology in 5G) on resource-constrained devices, and a proof-of-concept implementation as well. With this proof-of-concept platform, we have deployed a virtualised Voice over IP (VoIP) service. In both tests, we use Raspberry Pi 2 model B (RPi2)\footnote{https://www.raspberrypi.org/products/raspberry-pi-2-model-b/} boards. RPi2 are single-board computers, with a 900 MHz quad-core ARM Cortex A7 CPU, 1 GB RAM, 4 USB ports, 1 Ethernet port and a micro SD card slot. The election of this board is based on its capability to support virtualisation and, at the same time, a reduced cost, weight and low power consumption. Thus, it can be easily attached to almost any MAV to work independently or even integrate them to have a unified device to exchange information between the MAV controller and the RPi2. In all experiments, we use RPi2 boards with Raspbian, which is a Debian-based operating system optimised for our hardware.

\subsection{Viability of deploying virtual machines on resource-constrained devices}
We have defined a set of experiments to check the viability of using virtual machines (VMs) on RPi2 boards. The first experiment is designed to
find the maximum number of VMs supported by the board. After that, the following experiments compare the performance with and without VMs. For example, the second experiment presents the gap in performance
when an application is running on a VM against the native processing on the host. Finally, the third experiment shows the results of performance in terms of energy consumption.


In our first experiment,  VMs
should be configured to deploy a basic virtual service (a DNS server, for example) or a virtual network function (a proxy SIP). Thus, 
we have configured VMs with 1 CPU and 256 MB of RAM. We have used QEMU\footnote{http://www.qemu.org/} for the virtualisation functionalities and the KVM hypervisor\footnote{http://www.linux-kvm.org/} for the acceleration of the access to the low level hardware, which has support for the A7 ARM architecture. Moreover, all our VMs run a basic installation of OpenSuse for ARM. In order to increase the efficiency, one of the four available cores is isolated so to be dedicated to run the host operating system, and the three remaining cores are used for virtualisation purposes. With this configuration, the RPi2 can support up to four VMs, where the limit comes from the available total memory of 1 GB of RAM. In the scenario where four VMs
are running at the same time, two of them have to share the same core, while the remaining two VMs have dedicated cores. In the next experiment we want to check what is the reduction of performance of two or more VMs sharing a core, against 
a VM with a dedicated one. 


The measurement of the CPU consumption has been done by means of a stress application, which is a process that tries to use all the available CPU cycles, executing $10^5$ writing operations to /dev/null. The experiment consists on running one to four instances of this application on the hosts, and then repeating the same tests on one to four VMs. After the application executes all $10^5$ operations, it then
returns the total time consumed. Table\ref{tb:energyConsumption} collects the results obtained in this experiment when the processes are running on VMs (column 2) and when they are running on the host (column 3). In the last row, when 4 processes are running
at the same time, we present two results for VMs and two for the host. This is because two of those processes are running in
different virtual machines (1/VM) or CPUs (1/CPU), and the remaining 2 processes are running in the same virtual machine (2/VM) or CPU
(2/CPU). We can extract two results from this experiment: (1) there is a reduction of the performance when VMs are used, 
compared with the same application running on the host, and (2) the performance goes down when two processes are running
on the same VM, but this reduction is similar when these applications are running on the host. Although there is a lack of 
performance using VMs, we believe this is sometimes acceptable for some services. Furthermore, the gap can be reduced
using other techniques like containers or network stacks at the user level.


The next test is designed to check the energy consumption difference when VMs are used. The measurement of the energy consumption has been done using the wattmeter of the company Monsoon\footnote{https://www.msoon.com/LabEquipment/PowerMonitor/}. The consumption of the board in the steady state, with any additional process running apart of those used by the operating system, is 291mA, which is the value that will be use as reference for the rest of measures. When one VM is running, but no other process different that the operating system is running, the energy consumption is 293 mA. After this reference values are obtained, in this experiment we measure the energy consumed by the \emph{iperf}\footnote{https://iperf.fr/} application, which is used to
generate TCP or UDP traffic. In our case, \emph{iperf} is configured to transmit a UDP traffic of 400kbps towards a server running in another RPi2. There are eight measures: four with the energy consumption results when VMs are running and four when the \emph{iperf} processes are running on the host. The results of the experiment are shown in column \emph{Iperf} of table \ref{tb:energyConsumption}, where we can notice the increment in terms of energy consumption when VMs are used.

\begin{table}[ht]
\scriptsize
\caption{Energy consumption \& processing time}
\begin{tabular}{|M{3em}|M{10em}|M{10em}|M{2em}|M{2em}|}
\hline
 & \multicolumn{2}{c|}{\textbf{Stress}} & \multicolumn{2}{c|}{\textbf{Iperf}} \\ \hline
Process & VMs (s) & Host (s) & VMs (mA) & Host (mA) \\ \hline
1 & 19.73 & 6.1 &  345  & 296 \\ \hline
2 & 21 & 6.1 & 361 & 296 \\ \hline
3 & 23 & 6.15  & 372 & 296 \\ \hline
4 &  24(1/VM) and 40(2/VM)  & 6.3(1/CPU) and 11.3(2/CPU) & 382 & 298 \\ \hline
\end{tabular}
\label{tb:energyConsumption}
\end{table}

\subsection{Proof of concept}
\begin{figure}
\centering\includegraphics[width=1.0\columnwidth]{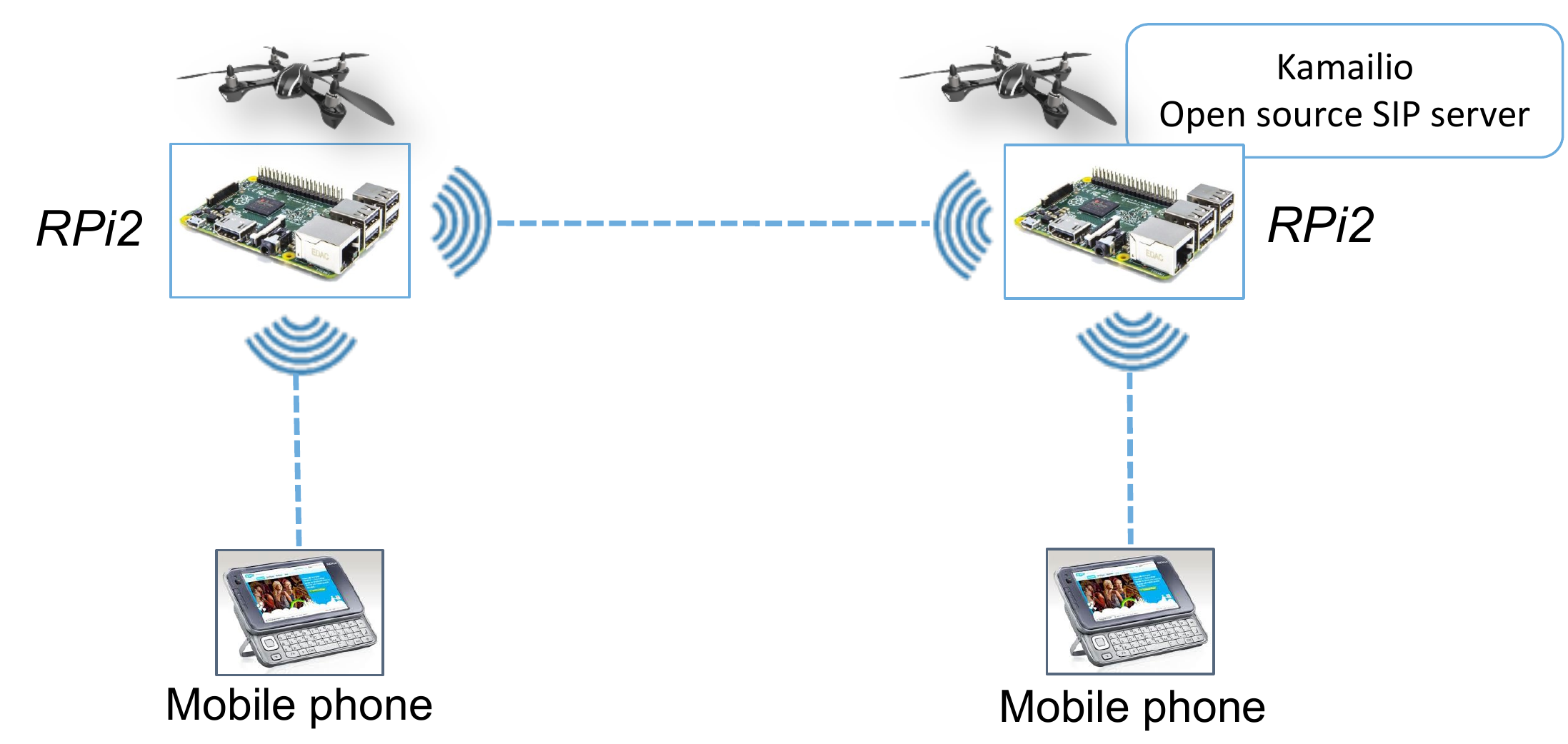}
\caption{Testbed used to deploy a simple VoIP service over MAVs}
\label{fig:proof-of-concept}
\end{figure}

The previous results show that virtualisation is feasible in resource-constrained platforms, such as those that can be available in MAV deployments. However, we have also seen that its use comes with a cost in terms of energy consumption and processing capacity. This may be a limiting factor in some scenarios. To complement our study of feasibility, we have carried out a practical experience considering a real use case. In this proof-of-concept, we evaluate the use of limited capacity platforms to enable voice/video communications in emergency situations (scenario B in section \ref{sec:use-cases}). For this purpose, we built up the simple testbed illustrated in Fig.~\ref{fig:proof-of-concept}, where two MAVs (represented as RPi2 boards) deploy the 5G functions that are necessary to allow the exchange of signalling and media between two mobile phones Nokia N810 (this model support the establishment of audio/video calls using SIP signalling and RTP). In our testbed, each RPi2 provides a WiFi access point, which is utilised by a mobile phone to get network connectivity. Both RPi2 are interconnected with a point-to-point WiFi link. One of RPi2 boards is used to deploy a SIP network function (by virtualising an instantiation of a open source SIP server\footnote{Kamailio: the Open Source SIP server. http://www.kamailio.org/w/}), to support the registration of mobile phones and the call control procedures.

\begin{figure}[ht]
\centering\includegraphics[width=1\columnwidth]{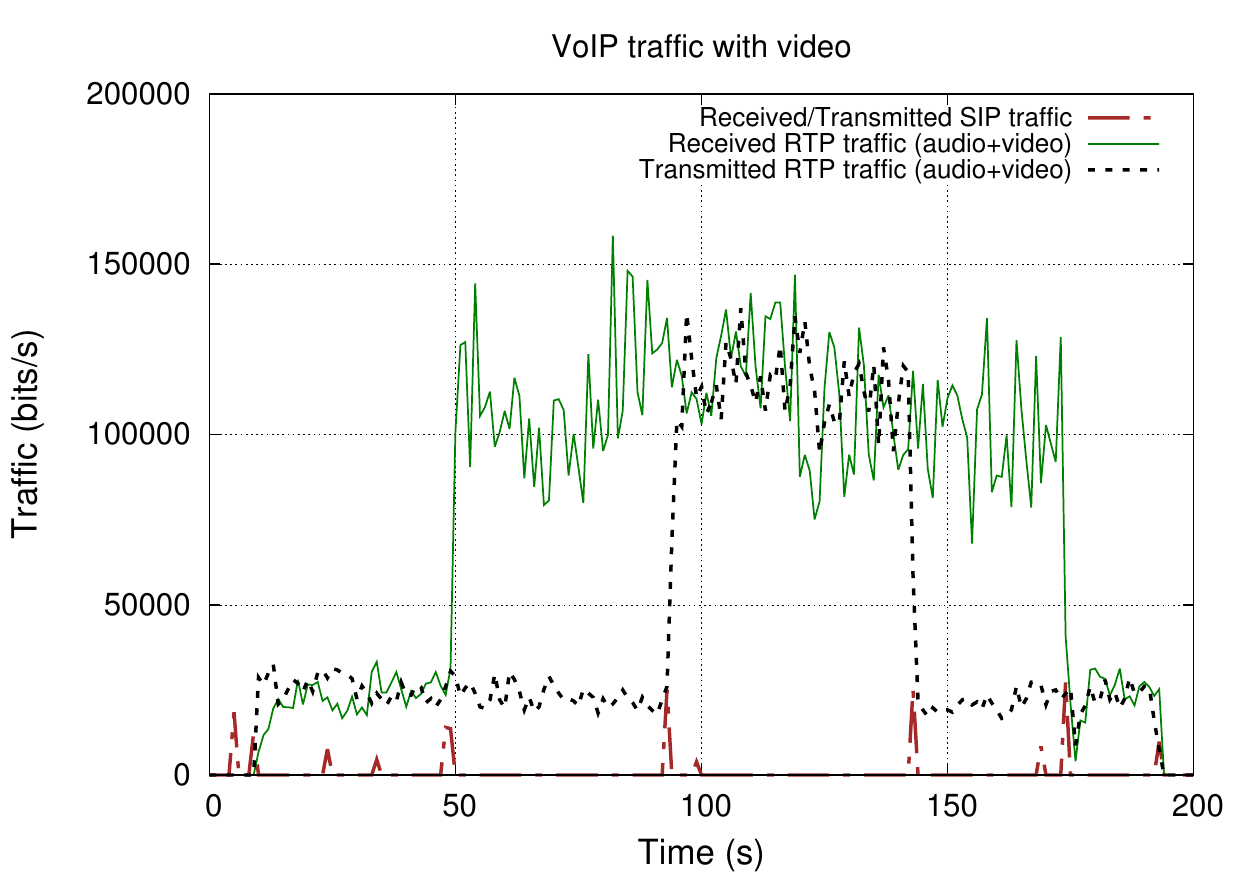}
\caption{Data exchange of a video call}
\label{fig:video-call}
\end{figure}

With this configuration, we performed several audio/video calls between both mobile phones. Figure \ref{fig:video-call} represents the data exchange corresponding to a specific call (captured in the access link of the user originating the call). In this case, the call starts with both users exchanging audio. After approximately 50 seconds, the terminating user activates the video camera of the mobile phone, and video starts being received by originating user. The video call proceeds with the originating user activating its own video camera (approximately at sec. 90). Finally, both users deactivate their video cameras and the call is terminated. The average energy consumption, in the RPI2 board executing the SIP server, was measured to be 437.92 mA during the audio exchange, increasing to 439.02 mA when both terminals activate the video delivery. We want to highlight that all the video calls established during the experiments performed with appropriate quality, with no voice/video glitches, which validates the feasibility of using virtualisation over MAV platforms in a real use case.

\section{Conclusion}
\label{sec:conclusion}
This paper explores the use of RPAs, and particularly MAVs, as 5G points of presence. As a first step, we evaluate the viability of using virtualisation technologies over resource-constrained devices, which may be the common baseline platform in MAV deployments. Our evaluation indicates that virtualisation is feasible in these emergent platforms, although the cost in terms of energy consumption may be a limiting factor in some scenarios. Our future work will continue the research work initiated in this paper, addressing a detailed analysis, both from a theoretical and practical perspective, of the advantages and limitations of RPAS platforms to support 5G functions in the use cases under consideration.

\section*{Acknowledgment}
The work in this article has been partially supported by the European H2020 Flex5Gware project (grant agreement 671563) and by the Spanish DRONEXT project (grant agreement TEC2014-58964-C2-1-R).

\end{document}